\documentclass[12pt]{article} 

\usepackage{amssymb,amsmath}
\usepackage[english]{babel}
\usepackage{graphicx}
\usepackage{url}

\begin{document}

\title{Mapping the narrow-line Seyfert 1 galaxy 1H~0323+342}

\author{Luigi Foschini\footnote{INAF Brera Astronomical Observatory, 23807 Merate, Italy.}\, \footnote{Correspondence: luigi.foschini@inaf.it; Tel.: +39-02-72320-458.}\, , Stefano Ciroi\footnote{Department of Physics and Astronomy, University of Padova, Italy.}, Marco Berton\footnote{Finnish Centre for Astronomy with ESO (FINCA), Turku, Finland.}\, \footnote{Mets\"ahovi Radio Observatory, Aalto University, Kylm\"al\"a, Finland.}\, ,\\Stefano Vercellone$^{*}$, Patrizia Romano$^{*}$ and Valentina Braito$^{*}$}
\date{January 29, 2026}
\maketitle

\begin{abstract}
Taking advantage of the most recent measurements by means of high-resolution radio observations and other multiwavelength campaigns, it is possible to elaborate a detailed map of the narrow-line Seyfert 1 Galaxy 1H~$0323+342$. This map will open the possibility of intriguing hypotheses about the generation of high-energy $\gamma$ rays in the narrow-line region.
\end{abstract}

\section{Introduction}
1H~$0323+342$, a.k.a. J$0324+3410$ ($z=0.063$\footnote{This measurement was done by \cite{ZHOU}, but an older value of $z=0.061$ was often used (e.g. NASA/IPAC Extragalactic Database). The differences are minimal: $\sim 7$\% for the luminosity, $1.21$~pc/mas vs $1.18$~pc/mas for the conversion of angle to linear size. In this work, we used all the quantities for $z=0.063$. In case we used values from literature with $z=0.061$, we corrected for the different $z$.}) is an active galactic nucleus (AGN) classified as jetted narrow-line Seyfert 1 Galaxy (jNLS1) \cite{ZHOU}, with a significant detection at high-energy $\gamma$ rays \cite{ABDO}. It is the closest source of this type and, therefore, it offers a unique opportunity to dissect the AGN, as one milliarcsecond (mas) corresponds to $\sim 1.2$~pc in a $\Lambda$CDM cosmology with $H_{0}=70$~km/s/Mpc, $\Omega_{\Lambda}=0.7$, $\Omega_{\rm{m}}=0.3$. This means that the Very Long Baseline Array (VLBA) can probe the AGN on a parsec scale. Indeed, J$0324+3410$ has been the target of many studies at high-resolution radio frequencies \cite{WAJIMA,FUHRMANN,DOI,HADA,KOVALEV}, and is currently under monitoring in the framework of the MOJAVE program\footnote{\textbf{M}onitoring \textbf{O}f \textbf{J}ets in \textbf{A}ctive galactic nuclei with \textbf{V}LBA \textbf{E}xperiments: \url{https://www.physics.purdue.edu/MOJAVE/}.} \cite{LISTER19}. It was found that the jet from J$0324+3410$ is one sided with a change in shape at about $\sim 7$~mas from the core, being parabolic close to the central black hole, and becoming conical as the distance increases \cite{DOI,HADA,KOVALEV}. This quasi-stationary feature is similar to HST-1 in M87, although the radial distance from the central singularity is $\sim 10^{7-8}r_{\rm g}$, two orders of magnitudes farther than HST-1 from M87 \cite{DOI,HADA,KOVALEV}. Both Hada \cite{HADA} and Kovalev \cite{KOVALEV} proposed that one solution could be to have a greater mass for the central black hole in J$0324+3410$, of the order of $10^{8}M_{\odot}$ as suggested by \cite{LEONTAV}.

The aim of the present work is to revise all the available information about the central mass black hole of J$0324+3410$, its surroundings and its jet, in order to obtain a consistent picture of this source.

\section{Mass of the central black hole}
The estimate of the mass of the central black hole in narrow-line Seyfert 1 Galaxies is at the center of a long-standing debate (see \cite{PETERSON} and references therein for the latest information). In the case of J$0324+3410$, the values are of the order of $10^{7}M_{\odot}$ (\cite{ZHOU,ABDO,FOSCHINI,WANG,LANDT}), with two exceptions \cite{LEONTAV,PAN}. The available values and adopted methods are summarised in Table~\ref{tab1}.

\begin{table}[h]
\caption{Summary of the mass estimates of the central black hole in J$0324+3410$.}
\centering
\footnotesize
\begin{tabular}{llc}
\hline
Mass & Method & Reference\\
($10^{7}M_{\odot}$) & {} & {}\\
\hline
$\sim 1$ & H$\beta$ luminosity and FWHM, single epoch spectrum & \cite{ZHOU}\\
$\sim 1$ & fit of the accretion disk, Shakura-Sunyaev model    & \cite{ABDO}\\
$\sim 3.6$ & H$\beta$ luminosity and $\sigma$, single epoch spectrum & \cite{FOSCHINI}\\
$3.4_{-0.6}^{+0.9}$ & Reverberation mapping H$\beta$ & \cite{WANG}\\
$2.2_{-0.6}^{+0.8}$ & single epoch spectrum, Pa$\alpha$, H$\alpha$, H$\beta$, excess variance at X-rays & \cite{LANDT}\\
$0.28-0.79$ & excess variance, PSD bend frequency at X-rays & \cite{PAN}\\
$16-40$ & Black hole-bulge relationship, $R$ and $K_{s}$ filters & \cite{LEONTAV}\\
\hline
\end{tabular}
\normalsize
\label{tab1}
\end{table}

\begin{figure}[h]
\centering
\includegraphics[angle=270,scale=0.5]{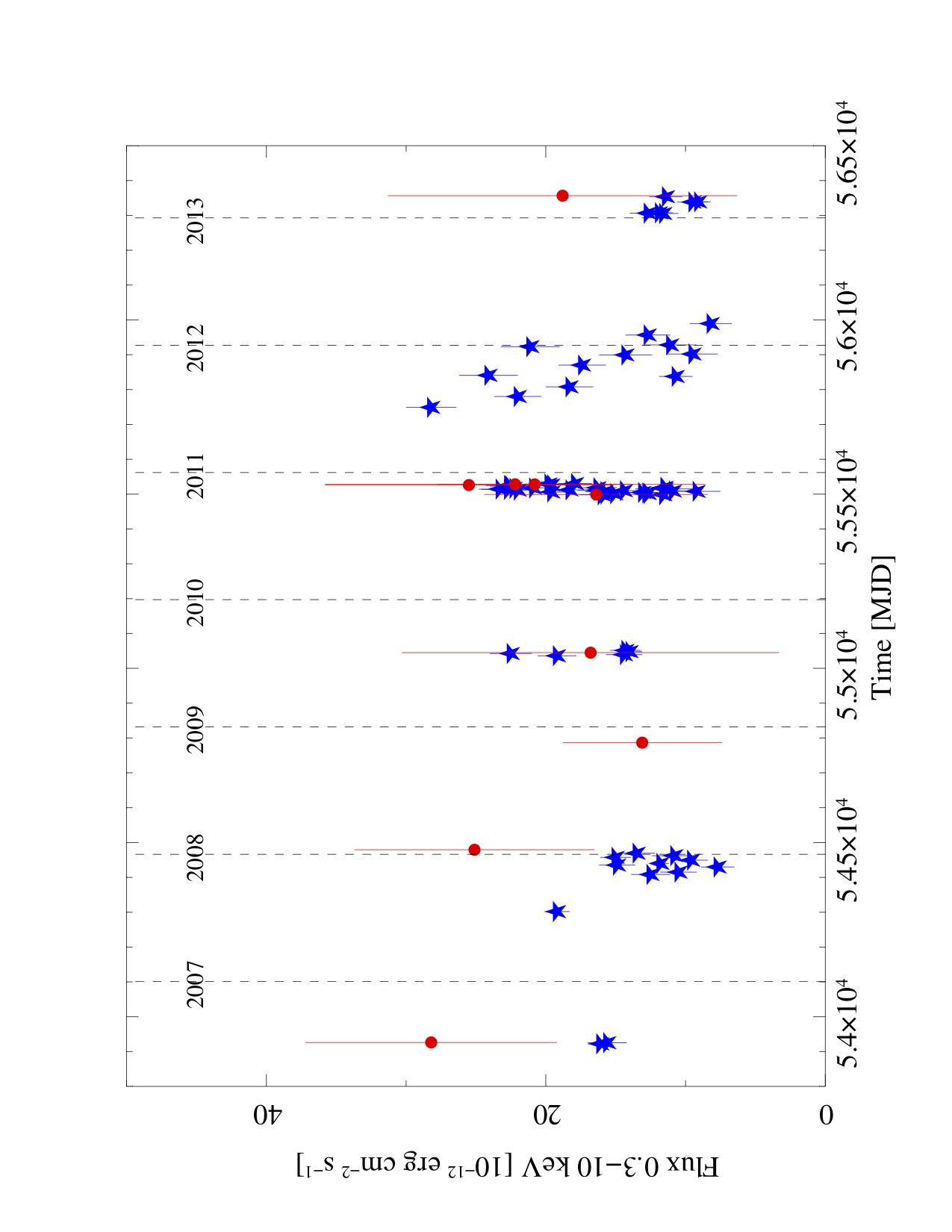}
\caption{\emph{Swift}/XRT light curve of J$0324+3410$ from the analysis reported by \cite{FOSCHINI}. Blue stars indicate a best fit with a power-law model; red circles are for a broken power-law model. The vertical lines for the years are placed on January 1st of each year.}
\label{fig1}
\end{figure}

The mass estimates through the fit of an accretion disk model is generally overestimated (e.g. Fig. 6 in \cite{GHISELLINI}), particularly when the jet contribution at infrared-to-ultraviolet frequencies is significant (e.g. see the case of J$0948+0022$, Fig. 6 in \cite{FOSCHINI}). However, this seems not to be the case for J$0324+3410$, as the disk is quite strong and often prominent over the continuum of the jet (cf. the spectral energy distribution in \cite{ABDO}). This is confirmed by the polarisation measurements at optical wavelength, which are quite low ($0.7-0.8$\% in $V$ filter, \cite{IKEJIRI}), and the X-ray emission ($0.3-10$~keV) is generally consistent with a thermal Comptonization (photon index $\sim 2$) from the corona. Only during the jet activity, there is a hard tail (photon index $\sim 1.4$) with a break at a few keV (see Fig. 1 in \cite{FOSCHINI12}; see also \cite{FOSCHINI09}). A rough estimate of the duty cycle of the jet can be done by selecting the X-ray observations where the broken power-law model was the best fit (jet active) and to compare with all the other observations best fitted by a power-law model consistent with the unsaturated Comptonization. From the data reported by \cite{FOSCHINI}, the global observing time was $208.5$~ks, and $34.8$~ks ($\sim 17$\%) of data resulted to be best fitted by a broken power-law model (Fig.~\ref{fig1}). This value of $\sim 17$\% could be taken as a rough estimate of the jet duty cycle. Therefore, it is not surprising to find an agreement between the estimate via accretion disk fit and that by using the reverberation mapping (RM). 

The RM measurement is also in agreement with the second order moment ($\sigma$) of the H$\beta$ emission line measured with one epoch spectrum. This quantity ($\sigma$) is known to be less affected than the full-width half maximum (FWHM) by the inclination and shape of the broad-line region (BLR), and the accretion rate of the disk (e.g. \cite{PETERSON} and references therein). Therefore, we can also exclude a bias due to these factors. 

In this work, we will consider as reference mass the average of the above cited values ($\sim 2.2\times 10^7 M_{\odot}$), by excluding the two outliers \cite{LEONTAV,PAN}. One reason for the result in excess by \cite{LEONTAV} could be the difficulty to remove the background, to model the point spread function (\cite{LEONTAV} noted the lack of high S/N stars in the field of view), and also the presence of evident signatures of a recent merger, discovered by \cite{ANTON}. The discrepancies in the mass estimate via X-ray variability \cite{LANDT,PAN} could be due to the level of activity of the jet during X-ray observations: as previously noted (Fig.~\ref{fig1}, see also \cite{FOSCHINI09,FOSCHINI12}), the X-ray corona is dominating the $0.3-10$~keV flux and only when the jet is strongly active a hard tail above a few keV emerges.

Having set the most likely mass of the central black hole to the value of $2.2\times 10^{7}M_{\odot}$, it is possible to derive the structure of the AGN. The gravitational radius is:

\begin{equation}
r_{\rm g} = \frac{GM}{c^2} \sim 3.2\times 10^{12} \, \mathrm{cm}.
\end{equation}

The size of the BLR is about $15$ light days ($3.9\times 10^{16}$~cm or $0.013$~pc), as derived from the measurements of H$\beta$ and Fe II reverberation mapping ($15_{-3}^{+4}$ and $15_{-4}^{+7}$ light days, respectively) \cite{WANG}. This size is consistent with the expected external radius of the accretion disk ($\sim 10^{4}r_{\rm {g}}$, \cite{KAWAGUCHI}). 
From the radius-luminosity relationship \cite{BENTZ} properly rearranged, it is possible to estimate the luminosity of the accretion disk:

\begin{equation}
L_{\rm disk} = \kappa(1.4\times 10^{-3}) R_{\rm BLR}^{1.88}\times 10^{44} \sim 2.3 \times 10^{44} \, \mathrm{erg/s}
\end{equation}

where $L_{\rm disk}$ is the disk luminosity, $R_{\rm BLR}=15$~light days, and $\kappa$ is the factor to convert the $\lambda L_{5100\AA}$ into the bolometric luminosity. This value is generally assumed to be $10$, although in case of super-Eddington accretors, it could be as great as $150$ \cite{CASTELLOM,NETZER}. In the above calculation, we assumed the conservative value of $\kappa=10$. In the case $\kappa=150$, then $L_{\rm disk}\sim 3.5\times 10^{45}$~erg/s.

The boundary between the BLR and the molecular torus is given by the dust sublimation radius \cite{ELITZUR}:

\begin{equation}
R_{\rm d,sub} = 0.4\sqrt{L_{\rm disk,45}} \sim 0.19 \, \mathrm{pc}
\end{equation}

where $L_{\rm disk,45}$ is the disk luminosity in units of $10^{45}$~erg/s. The dust sublimation temperature has been taken to be $1500$~K, and different values imply an additional factor \cite{ELITZUR}. The outer boundary of the torus is given by \cite{ELITZUR}:

\begin{equation}
R_{\rm out} = 12\times \sqrt{L_{\rm disk,45}} \sim 5.7 \, \mathrm{pc}.
\end{equation}

The extension of the narrow-line region (NLR) can be measured from the luminosity of the [O III]$\lambda$5007 line emission \cite{FISCHER}. This line has been carefully studied by \cite{BERTON}, who found that the profile was significantly affected by turbulence, likely due to the interaction with the relativistic jet. The line luminosity was found to be $\log L_{\rm [O III]}=40.98$ \cite{BERTON}, which corresponds to a maximum extension of the NLR \cite{FISCHER}:

\begin{equation}
R_{\rm NLR,max} = 10^{(-18.41 + 0.52\times \log L_{\rm [O III]})} \sim 794 \, \mathrm{pc}
\end{equation}

The ring/spiral arm structure observed by \cite{ZHOU,ANTON,LEONTAV} has a radius of $\sim 7.5''$, corresponding to a linear size of $\sim 9$~kpc. The viewing angle is unknown, but, by considering this structure developed along the equatorial plane of the central black hole, and that the jet is perpendicular to the black hole equator, then the viewing angle of the ring/spiral arm should be the complementary angle of the jet viewing angle $\theta\sim 9^{\circ}$ (see the next Section). Therefore, the deprojected size is almost equal to the above calculated linear size.

\section{The jet structure}
The structure of the relativistic jet from J$0324+3410$ has been studied at high-resolution radio frequencies \cite{WAJIMA,FUHRMANN,DOI,HADA,KOVALEV}. Here we summarise their results. The jet shows a break at a quasi-stationary bright feature located downstream at $\sim 7$~mas from the radio core. The inner region of the jet has a parabolic shape, while the outer region has a conical shape (see Fig. 2 in \cite{HADA}). At the collimation break, the jet width is squeezed to half its value (see Fig. 5 in \cite{DOI}). Taking as reference the radio core at $43$~GHz, and assuming that the jet has the origin close to the black hole position, then the jet width as a function of the radial distance from the black hole $W(r)$ [mas] is \cite{HADA}:

\begin{equation}
W(r) \propto \begin{cases}
(r-r_{0})^{0.60\pm0.03} & r \lesssim 7 \,\mathrm{mas}\\
(r-r_{0})^{1.41\pm0.02} & r \gtrsim 7\,\mathrm{mas}
\end{cases}
\end{equation}

where $r_{0}=-41\pm36$~$\mu$as is the distance of the radio core at $43$~GHz from the estimated position of the central black hole.

To calculate the linear distance, it is necessary to know the viewing angle of the jet. Many authors limited the viewing angle to $\theta \leq 12^{\circ}$ \cite{FUHRMANN,HADA}. Recently, by measuring the Doppler factor from the variability in $15$~GHz OVRO\footnote{Owens Valley Radio Observatory.} light curves ($\delta = 5.70_{-1.02}^{+1.01}$) and the $\beta_{\rm {app}}=9.05$ from the kinematic of the components as monitored by the MOJAVE project \cite{LISTER16,LISTER19}, a new value of $\theta = 9.06_{-8.91}^{+1.04}$~degrees has been calculated \cite{LIODAKIS}. The corresponding bulk Lorentz factor is then $\Gamma\sim 10$.

We can now convert the angular distances to linear size by taking into account the redshift $z=0.063$, which implies a projected linear size of $\sim 1.2$~pc/mas, and the jet viewing angle $\theta \sim 9^{\circ}$, which in turn means a deprojected size of $\sim 7.7$~pc/mas. A smaller viewing angle (says $4^{\circ}$, \cite{FUHRMANN} or even $3^{\circ}$, \cite{ABDO}), implies a greater deprojected distance ($\sim 23$~pc/mas in the case of $\theta \sim 3^{\circ}$). The results are summarised in Table~\ref{tab2}. 

\begin{table}[h]
\caption{Conversion of angular sizes to linear and deprojected sizes.}
\centering
\begin{tabular}{lcccc}
\hline
Feature & Angular & Linear & Deprojected & Reference\\
{} & (mas) & (pc) & (pc) & {}\\
\hline
$43$~GHz Core & $0.041$ & $0.050$ & 0.32 & \cite{HADA}\\
Convergence   & $6.97$  & $8.4$ & $54$ & \cite{DOI}\\
Intensity Peak   & $7.40$  & $9.0$   & $57$ & \cite{DOI}\\
Jet length $1.4$~GHz & $12.5\times 10^3$ & $15\times 10^3$ & $96\times 10^3$ & \cite{ANTON}\\
\hline
\end{tabular}
\label{tab2}
\end{table}%

Asada \& Nakamura \cite{ASADA} suggested that the change in shape of the M87 jet is located close to the Bondi radius. In the case of 1H~$0323+342$, that quantity is \cite{ALLEN}:

\begin{equation}
r_{\rm Bondi} = \frac{2GM}{c_{\rm s}^2} \sim 2.6\times 10^{18}\, \mathrm{cm} \sim 0.84\, \mathrm{pc}
\end{equation}

where $c_{\rm s}$ is the adiabatic sound speed of the X-ray emitting gas at the accretion radius $r_{\rm Bondi}$. If we assume a gas temperature similar to that of M87 ($kT=0.8$~keV, \cite{ALLEN}) and an adiabatic index $\gamma=5/3$, then the sound speed is $c_{\rm s}\sim 454$~km/s. The result is a radius consistent to the case of M87: in terms of gravitational radius, we have $\sim 8.1\times 10^{5}r_{\rm g}$ for 1H~$0323+342$ vs $\sim 7.6\times 10^{5}r_{\rm g}$ for M87 \cite{ASADA}.

\section{The map}
It is now possible to summarise the linear deprojected size of all the structures of the jNLS1 J$0324+3410$ in terms of gravitational radius, thus mapping this cosmic source. Table~\ref{tab3} lists all the structures of the AGN starting from the central black hole to the outer part of the galaxy, while Fig.~\ref{fig1} displays the corresponding schematic.  

\begin{table}[h]
\caption{Map of the narrow-line Seyfert 1 Galaxy J$0324+3410$. See Fig.~\ref{fig2} for a schematic.}
\centering
\begin{tabular}{lcc}
\hline
Feature & Size & Size \\
{}      & ($r_{\rm g}$) & (pc)\\
\hline
Gravitational Radius & 1 & $1.0\times 10^{-6}$\\
\hline
Broad-line region    & $1.3\times 10^4$ & $1.3\times 10^{-2}$\\
Torus (inner radius) & $1.9\times 10^5$ & $0.19$\\
\hline
$\>$ Jet $43$~GHz Core        & $3.2\times 10^5$ & $0.32$\\
\hline
Bondi radius         & $8.1\times 10^5$ & $0.84$\\
\hline
Torus (outer radius) & $5.7\times 10^6$ & $5.7$\\
\hline
$\>$ Jet Convergence (nozzle)  & $5.4\times 10^7$ & $54$\\
$\>$ Jet Intensity Peak       & $5.7\times 10^7$ & $57$\\
\hline
Narrow-line region (outer radius) & $7.9\times 10^8$ & $794$\\
Ring/Spiral Arm      &  $9.0\times 10^9$  & $9\times 10^3$\\
\hline
$\>$ Jet length      &  $9.6\times 10^{10}$  & $9.6\times 10^4$\\
\hline
\end{tabular}
\label{tab3}
\end{table}%

\begin{figure}[h]
\centering
\includegraphics[scale=0.3]{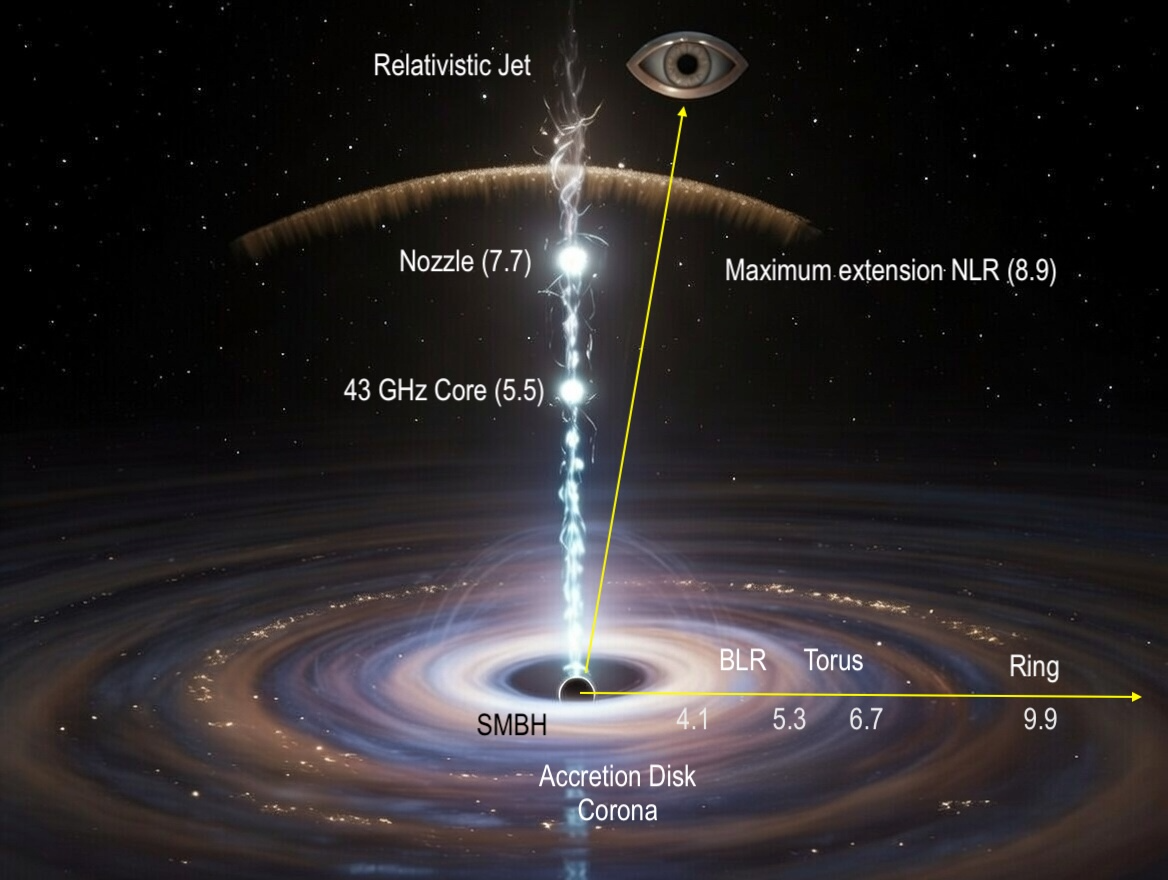}
\caption{Schematic of the AGN J$0324+3410$ elaborated by Grok, created by xAI. The distances (explicitly indicated with numbers) are in units of gravitational radius and in logarithmic scale (cf. Table~\ref{tab3}).}
\label{fig2}
\end{figure}

As noted earlier, some authors \cite{HADA,KOVALEV} suggested that a central black hole with a greater mass, consistent with the measurements by \cite{LEONTAV}, would imply the overlap of the jet profile of J$0324+3410$ with that of M87. However, this line of reasoning is the classical \emph{post hoc, ergo propter hoc}. Indeed, the scalability of the jet is a \emph{consequence} of the self-similar shape, as proven by \cite{HEINZ}. High-resolution radio observations of M87 and J$0324+3410$ proved that the jet shape is self-similar (paraboloidal and conical) in both sources. Therefore, it is possible to scale the jet according to \cite{HEINZ}, but it is not useful to set the mass of the central singularity, because it is the normalisation quantity. This also means that we can make the two jet profiles overlap by setting a smaller mass for the black hole in M87. 

We note from Table~\ref{tab3} that the $43$~GHz core is placed at a distance comparable with that of the torus\footnote{Obviously, it is not \emph{on} the torus, as the two structures are perpendicular each other.}. It is worth noting that, in case of super-Eddington accretion, the greater disk luminosity ($L_{\rm disk}\sim 3.5\times 10^{45}$~erg/s), would imply a farther torus ($R_{\rm d,sub}\sim 0.75$~pc), much farther than the radio core. Moreover, if we assume the upper limit of the BLR ($15+7=22$ light days) and a super-Eddington disk ($\kappa=150$), then $L_{\rm disk}\sim 7\times 10^{45}$~erg/s, which in turn implies $R_{\rm d,sub}\sim 1$~pc and $R_{\rm out}\sim 32$~pc. Therefore, in this case, the radio core would be still at the scale of the BLR, but the jet nozzle remains always outside the torus, whatever the combination of measurements and errors we could consider. 

The quasi-stationary feature at $\sim 5\times 10^{7}r_{\rm g}$, which is clearly in the NLR, is much more interesting. On one side, this is not surprising: the [O III] emission line profile reveals features indicating high turbulence, likely due to the interaction of the jet with the NLR gas \cite{BERTON}. On the other side, this is surprising when we linked this information with the observed activity at high-energy $\gamma$ rays \cite{DOI,HADA}. The jet width at the convergence is $0.39\pm 0.17$~mas \cite{DOI}, corresponding to $\sim 0.47$~pc (or $4.5\times 10^{5}r_{\rm g}$). The fastest variability at high-energy $\gamma$ rays was measured to be $\sim 3.1$~hours \cite{PALIYA}, which implies a size of the emitting region $r<\tau c \delta /(1+z)\sim 1.8\times 10^{15}$~cm. This is clearly too small for the jet size at the convergence shock. The observed width of the jet is more consistent with a time scale of months ($\sim 97$~days), and invoking a greater, but still reasonable, Doppler factor is not sufficient to mitigate the requirements (for $\delta=50$, the time scale could be $11$~days). A dedicated multiwavelength campaign with an intense sampling (hours time scale) would be necessary to better understand the behaviour of the quasi-stationary shock and the possible production of high-energy $\gamma$-rays. 

If confirmed, it would open a really intriguing possibility. The debate about the site of $\gamma$-ray generation was always confined between the BLR \cite{GHISELLINI14}, the torus \cite{MARSCHER,SIKORA}, or both cases, depending on the epoch, as in the case of PKS~$1222+216$ \cite{FOSCHINI11bis}. Perhaps, now it is time to add one more competitor: the NLR. In this case, being the seed photons co-spatial with the relativistic electrons of the jet as shown by the turbulence in the [O III] line emission \cite{BERTON}, it is necessary to study the new dynamics of interactions in order to understand the possible implications.

\section*{Acknowledgments} We acknowledge financial contribution from the agreement ASI-INAF n. 2017-14-H.0.



\end{document}